# A NEUTRON TRANSMISSION STUDY OF ENVIRONMENTAL GD

CRISTIANA OPREA[1], IOAN ALEXANDRU OPREA[1], ALEXANDRU MIHUL[2]

[1]*Joint Institute for Nuclear Research (JINR), Frank Laboratory for Neutron Physics (FLNP), Dubna 141980, Russian Federation*

[2]*University of Bucharest, Nuclear Physics Department, Bucharest-Magurele, RO-077125, Romania*

*Abstract.* A new method for the determination of environmental Gd by neutron transmission (NT) experiments is proposed. The NT method is based on the measurements of neutron spectra passing through a target. From the attenuation neutron spectra new data as concentration, width, resonance energies and cross section have been obtained.
By a computer simulated experiment in a common sample of 6 elements including $^{160}$Gd nucleus the concentration with an acceptable error lower than 10% was evaluated. Some other cases have been analyzed in order to improve the measurements. It was concluded that a combination of few analytical methods works well for the stable isotopes which are difficult to be determined by conventional neutron activation methods.

## 1. INTRODUCTION

Instrumental Neutron Activation Analysis (INAA) is a powerful method for the identification of nuclei in different type of samples. Unfortunately the presence of many environmental elements cannot be determined by INAA methods and this is mainly caused by the fact that by neutron capture stable isotopes are produced and not radioactive nuclides [1]. Between the elements which cannot be evaluated by INAA are several light elements and some medium and heavy nuclei. The authors of the reference [1] suggested the Prompt Gamma Activation Analysis (PGAA) or their own new methods based on the decreasing of neutron flux in the presence of studied elements and/or an indicator sample for the determination of named elements. They have defined their method as a new type of INAA and they named this new approach as Instrumental Neutron Absorption Activation Analysis (INAAA).

The authors of the present work have earlier proposed a method for the evaluation of the concentration and other properties of the light nuclei $^6$Li and $^{10}$B consisting in a combination of INAA and NT experiments [2]. Both nuclei have a very low value of neutron capture cross-section and a very high value of (n,α) rate for thermal neutrons [3]. These values for $^6$Li are:

$$\sigma_{n\alpha}(E_{th}) = (940 \pm 4)b, \quad \sigma_{n\gamma}(E_{th}) = (0.0385 \pm 0.003)b \tag{1}$$

and for $^{10}$B:

$$\sigma_{n\alpha}(E_{th}) = (3837 \pm 9)b, \quad \sigma_{n\gamma}(E_{th}) = (0.5 \pm 0.2)b \tag{2}$$

The very high value of (n,α) cross section gives the possibility to observe these element in the neutron spectra obtained from NT measurements.

The Gd nucleus has many stable isotopes [4,5] as follows: $^{152}$Gd (0.20%), $^{154}$Gd (2.18%), $^{155}$Gd (14.80%), $^{157}$Gd (15.65%), $^{158}$Gd (24.84%) and $^{160}$Gd (21.86%) [6]. From them only $^{160}$Gd isotope produces by neutron capture the radionuclide $^{161}$Gd with the time of life approximately three minutes [3] which is also difficult to measure by NAA. The Gd nucleus is metal from the III$^{rd}$ group of Medeleev periodical Table of Elements and belongs to Lanthanides. This metal is important in applications connected to the field of modern nuclear energetics and electronics.

In this study the possibility to evidence the presence of Gd nucleus by NT measurements through a computer simulation is performed.

## 2. NEUTRON CAPTURE

In the interaction of slow neutrons with nuclei a compound nucleus which is decaying on energetic possible channels is formed. The formed compound nucleus after the neutron capture is excited, has the same properties namely spin, parity, mass, etc, as any other stable nucleus and can be described by quantum states or resonances. The number of neutron resonances increases with the mass of compound nucleus. In the case of well isolated resonance the cross section of the interaction of neutrons with nuclei has the well known Breit – Wigner form [7]:

$$\sigma_{nx} = \pi \lambda^2 \frac{(2J+1)}{(2I+1)(2s+1)} \cdot \frac{\Gamma_n \Gamma_x}{(E - E_{rez})^2 + \frac{\Gamma_{tot}^2}{4}} \tag{3}$$

where: *J, I, s* = spin of compound nucleus, target nucleus and neutron; $\Gamma_n, \Gamma_x$ = neutron, x-particle widths; $\Gamma_{tot} = \Gamma_n + \Gamma_{n'} + \Gamma_\gamma + \Gamma_p + \Gamma_\alpha + \Gamma_d + \Gamma_t + \Gamma_{He^3} + \Gamma_f + ... =$ total width; $x = n, n', p, d, t, He^3, \alpha, f$ + other nuclear clusters, $E_{rez}$ = resonance energy; $\lambdabar = (2\pi)^{-1}\lambda = k^{-1}$ = reduced neutron wavelength;

For a better description and comparison of cross sections with experimental data generally the neutron resonance parameters and widths from [3] are used since they are obtained experimentally. In the case of slow neutron interaction usually are considered only neutrons with orbital momentum l = 0 (s neutrons) and l = 1 (p neutrons). If s and respectively p neutrons interact with the target nucleus S and P resonances of compound nucleus are formed. The parametric relations for s and p neutron widths, respectively, according with [3] are:

$$\Gamma_n^S(E_n[eV]) = \Gamma_{n0}^S \sqrt{E[eV]} \tag{4}$$

$$\Gamma_n^P(E_n[eV]) = \Gamma_{n1}^P \sqrt{E_n[eV]} p_1(E_n) \tag{5}$$

$$p_1(E_n) = (kR)^2 \cdot [1 + (kR)^2]^{-1} \tag{6}$$

where $\Gamma_{n0}^S, \Gamma_{n1}^P$ = reduced neutron widths for s and p neutrons; $p_1(E_n)$ = suppression factor; $k$ = neutron length number, $R[fm] = 1.45 A^{1/3}$ = nucleus radius.

The suppression factor is a result of the increasing of nuclear barrier height by centrifugal component which reduces the probability of the neutron to pass through barrier and therefore S resonances are large and P resonances are thin. The energetic dependences of S and P resonances are represented in Figure 1 based on relations (3) - (6).

In the real spectra more than one resonances yield but for slow neutrons these resonances can be considered isolated, well defined and far away one from each other. This condition can be written as:

$$\Gamma_i \langle\langle |E_i - E_j| \tag{7}$$

In this case the cross section is a sum of contributions given by the each resonance:

$$\sigma_{nx} = \sum_{i=1}^{N_r} \sigma_{nx}^i \tag{8}$$

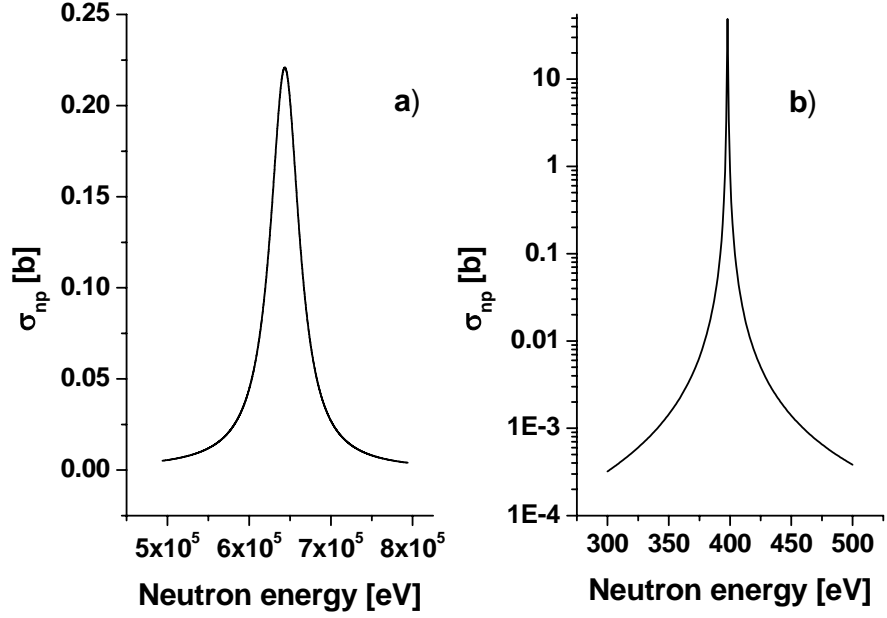

Figure 1. Cross sections calculated with parameters from [3] and formulas (3) - (6) - a) Large S-resonance with energy $E_S$ = 648 keV in $^{14}N(n,p)^{14}C$ reaction;   b) Thin P-resonance with energy $E_P$ = 398 eV in $^{35}Cl(n,p)^{35}S$ reaction.

## 3. NEUTRON TRANSMISSION

A powerful and efficient method in the investigation of properties of nuclei as concentration, cross section, widths, resonance energies, spin, parity and others is the NT method. In this method the new data on the studied nuclei are extracted from neutron spectra obtained from attenuated neutron flux passing trough a target. The attenuation of neutron flux has the expression [8]:

$$T = Exp\left[-\sum_{i=1}^{nr\_elem} n_i \sigma_{tot}^i\right] \quad (9)$$

where $n_i$ = concentration of i-th element from the target [$m^{-2}$], $\sigma_{tot}^i$ = total cross section of i-th element and *nr_elem* = number of elements from the target.

For slow neutrons only the S resonances gives the main contribution in the cross section and they were further considered in the present evaluation. Than for the i$^{th}$ element the total cross section has the form:

$$\sigma_{tot}^{i} = \sigma_{pot}^{i} + \sigma_{rez}^{i} + \sigma_{pot+rez}^{i-\text{int}\,erf} =$$

$$= 4\pi R_i^2 + \sum_x \left( g_{Si}^x \pi \lambda^2 \frac{\Gamma_n^{Si}\Gamma_x^{Si}}{\left(E - E_{Si}^x\right)^2 + \left(\frac{\Gamma_{tot}^{Si}}{2}\right)^2} \right) + \frac{4 g_{Si}^x \pi \lambda \Gamma_n^{Si} R_i (E - E_{Si})}{\left(E - E_{Si}\right)^2 + \left(\frac{\Gamma_{tot}^{Si}}{2}\right)^2} \quad (10)$$

where $R_i = 1.45 A_i^{\frac{1}{3}} [fm] = $ radius of i-th nucleus and $A_i = $ mass number.

The first term in (10) is the potential scattering. The second term represents the resonant interaction on different open channels ($x=n, p, d, t, {}^3He, \alpha,$ etc.) and the third term comes from the interference between potential and resonant neutron scattering.

In many cases the interference between resonant and potential scattering can be neglected. For slow neutrons where only neutron and capture channels are open the resonant – potential scattering interference can be observed because gives a deviation from the known *1/v* law of cross section.

From the above relations it is easy to understand that the determination of the properties of studied nuclei is based on the presence of resonances in the spectra. One of the most important elements in the data processing is the broadening of the resonance due to the thermal motion of target nuclei (Doeppler resonance broadening). As result the value of the cross section in the resonance, consistently in many cases, is decreasing by going far from the resonance. This effect must be taken into account in the cross section and widths evaluation. Also this effect plays a crucial role in construction, maintenance and well functioning of nuclear energetic installations [7]. Considering the approximation of gas target [8] and the Maxwell – Boltzmann distribution of neutron velocities the effective cross section measured in the experiment is:

$$\sigma_{eff}(E_n) = \frac{1}{\pi\sqrt{\Delta}} \int_0^\infty \sigma(E) Exp\left[-\left(\frac{E - E_n}{\Delta}\right)^2\right] dE \quad (11)$$

with $\Delta = \sqrt{\frac{4 m_n E_n k_B T}{A}} = $ Doeppler width, where $m_n = $ neutron mass, $E_n = $ neutron energy, $k_B = $ Boltzmann constant, $T = $ temperature of the target in ${}^0$K and $A = $ mass of the target nucleus.

Expression (11) is used for the temperature correction in the cross section. To extract the real cross section it is necessary to solve the integral equation (11) or to use the data tables for temperature corrections. In Figure 2 Doppler temperature broadening is represented for the neutron resonance of ${}^{35}$Cl nucleus with the energy $E_S = 14.8$ keV for different temperatures. The main temperature effect is in the resonance and close to the resonance.

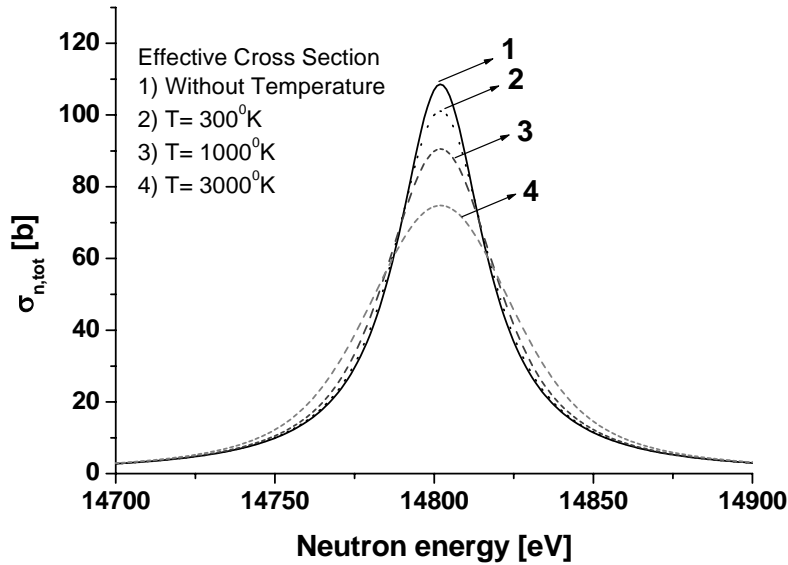

Figure 2. Temperature broadening of the neutron total cross section for $^{35}$Cl nucleus for S resonance, $E_S = 14.802$ keV, for different temperatures.

For many decades in the FLNP JINR Dubna groups using NT measurements in combination with neutron time of flight spectrometry are dealing, improving in this way the effectiveness of the NT method. Interesting applicative questions for nuclear and reactor technology connected with the determination of energies, spins, parities, widths of resonances for many elements were solved. Also interesting fundamental tasks as parity violation observed in slow neutron reactions were underlined. A comprehensive description of the NT method, time of flight neutron spectroscopy and resonance broadening used in the FLNP JINR Dubna experiments can be found in references [9-11].

Neutron transmission experiments are based mainly on the existence of neutron resonances of the compound nucleus which are formed in the interaction of slow neutrons with the target nuclei. The resonances of the compound nucleus in the case of slow neutrons can be very well evidenced using a pulsed neutron source and the time of flight spectroscopy with a high precision.

The setup of a neutron transmission experiment is simple as shown in Fig. 3.

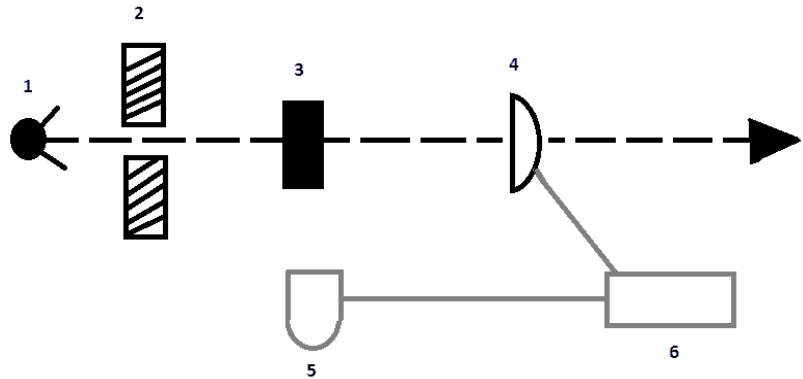

Figure 3. The experimental setup of NT experiment. 1- Pulsed neutron source; 2- Collimators; 3- Sample; 4- Neutron detector; 5- Gamma detector; 6- Other type of detectors

In the experimental setup, using time of flight method and the corresponding electronic devices, the neutron energy is determined by measuring the time necessary as the neutron passes from source to the sample. This time depends on the neutron energy and therefore from the attenuation of the incident flux the neutron spectrum is obtained. Since for slow neutrons the resolution of time of flight method is high the resonances can be observed in the spectra.

The neutron detectors in the NT methods are used for total cross section evaluation. For improvement of the measurements can be used gamma, fission and other type of detectors for partial cross section determination of the capture, fission and other processes.

In the analysis of environmental, industrial and other type of samples NT method can be applied as this method (like INAA) is a non distructive one and not necessitates a special preparation of samples before measurements. Also, this approach allows to investigate large samples with volume of order of tens cubical decimeters (of thickness up to some tens of centimeters).

## 4. COMPUTER SIMULATED EXPERIMENT

The neutron transmission can be a complementary method to INAA in the evaluation of properties of elements which are not possible to be determined by INAA. As suggested in [1] for the determination of light elements, Cd, Gd and other nuclei can be used PGAA, but this method is more difficult and expensive than NT. Furthermore gamma detectors can replace the neutron detectors in an NT experimental setup and then it is possible to obtain gamma spectra of elements from the sample [8, 9].

For the determination of $^{160}$Gd nucleus we realized a computer simulated experiment. In a sample which will be irradiated with neutron beam with energy up to 16 keV, it is supposed to have a number of six elements and between them it is the $^{160}$Gd nucleus with concentrations of order of ppm. From the beginning it is considered that the composition of the sample matrix is well known and determined by other methods. Simulated experimental data are generated using the relations (1) to (10) considering each point spreaded according to Gauss Law. An analogue software was realized by authors in a proposal for estimation of the presence of light nuclei [2]. With not major modifications of the software for light nuclei the main parts of the codes for $^{160}$Gd nucleus are:

1. Theoretical data simulation;
2. Simulated experimental data for NT for all open channels of elements in the sample ;
3. Least square methods for NT experiments;
4. Error evaluation of simulated experimental data;
5. Extraction of the necessary information;
6. Graphic representation section with export option on ACII files for other graphical tools.

## 5. RESULTS AND DISCUSSION

The simulated NT experiments were realized for the preparation of the measurements to obtaining nuclear data at IREN - the new basic facility neutron source of FLNP JINR Dubna. As in a real experiment the concentration of $^{160}$Gd was obtained. The following elements with one S resonance, respectively, with energy in eV: Mo - 44.7 ($E_1$), Cu - 579 ($E_2$), Fe - 7760 ($E_3$), Cl - 14802 ($E_4$), Sr – 3.54 ($E_5$), Gd – 904.6 ($E_6$) were chosen.

For simplicity only one S resonance for each element was taken into account. Following the cross section formulas (3, 5) and Figure 1 it can be observed that the S resonances give the main contribution to the cross section. In the future new evaluations it will be necessary to introduce all resonances for the all elements and to take into consideration possible interferences between resonances of the same element of overlapping between resonances of different nuclei.

One important source of error is the thermal motion of target nuclei and this error affects especially the resonances.

In Table 1 the corrections for each element and resonance energy calculated by relation (11) are given.

Table 1.
Cross section values due to the thermal motion for resonance energy at different temperatures for each element. In the second row the cross section is calculated according to (1)

| Sect[b]/T[$^0$K] | $E_1$ | $E_2$ | $E_3$ | $E_4$ | $E_5$ | $E_6$ |
|---|---|---|---|---|---|---|
| Without corr. | 18131 | 1991 | 334 | 108 | 1162 | 2792 |
| 300 | 12057 | 1419 | 333 | 101 | 1008 | 2583 |
| 1000 | 8616 | 1044 | 333 | 91 | 840 | 2295 |
| 3000 | 5819 | 720 | 333 | 75 | 642 | 1877 |

In a common chosen case the concentrations of the five elements is known from other measurements with a determined error. In our computer program we fixed the error for the first five elements of about 10-15%. Using them the concentration of $^{160}$Gd nucleus is extracted (Table 2). The all elements have the concentration of order of ppm ($10^{-6}$ mg/kg).

Table 2.
Concentrations of the first five elements were determined initially with a given error

| Conc. $10^{22}$ m$^{-2}$ | $E_1$ | $E_2$ | $E_3$ | $E_4$ | $E_5$ | $E_6$ |
|---|---|---|---|---|---|---|
| Initial | 6.4 ± 0.6 | 8.4 ± 0.8 | 8.75 ± 0.9 | 3.57 ± 0.46 | 1.85 ± 0.20 | 2.96 |
| Final |  |  |  |  |  | 3.40 ± 0.05 |

Another situation yields when the concentration of $^{160}$Gd is extracted simultaneously with the all elements from NT spectra. The results are given in Table 3. In the second row are the initial data for all elements and in the third row are the extracted concentrations.

Table 3.
Evaluation of Gd and other constituent concentrations from NT spectra

| Conc. $10^{22}$ m$^{-2}$ | E1 | E2 | E3 | E4 | E5 | E6 |
|---|---|---|---|---|---|---|
| Initial | 6.43 | 8.33 | 8.73 | 3.47 | 1.73 | 2.96 |
| Final | 5.96 ± 0.77 | 9.05 ± 0.06 | 8.9 ± 0.03 | 3.21 ± 0.49 | 1.72 ± 1 | 2.76 ± 0.05 |

The number of generated simulated data was 1024 in a wide energy interval up to 16 keV. The concentration of $^{160}$Gd nucleus was determined very well in both situations. Still from Table 3 it is observed that elements 1 and 5 are evaluated with relative big errors. These elements have close resonances and partially overlapping. The resonance of Gd nucleus is practically isolated and therefore in both cases the concentration is obtained with a very good precision. As it was expected if near the Gd resonance another overlapping resonance from other element is introduced the precision of determination is decreasing slower or faster depending on the degree of overlapping between resonances.

The evaluation of elemental concentrations is related to applicative researches. In fundamental researches many times the widths determination is of interest. Like in precedent case the NT spectra is generated from a sample composed by the same 6 elements including the $^{160}$Gd nucleus. It is supposed initially that the concentrations for all elements are known. The value of reduced neutron width is [3]:

$$\Gamma_n^0 = 0.1166 \, \text{eV} \tag{12}$$

Considering that the all elements have the concentration of the order of $10^{22}$ m$^{-2}$ the NT spectrum is generated (Figure 3). Then the new value of reduced neutron width is extracted by fit procedure as:

$$\Gamma_{n0}^{fit} = (0{,}14 \pm 0.03) \, \text{eV} \tag{13}$$

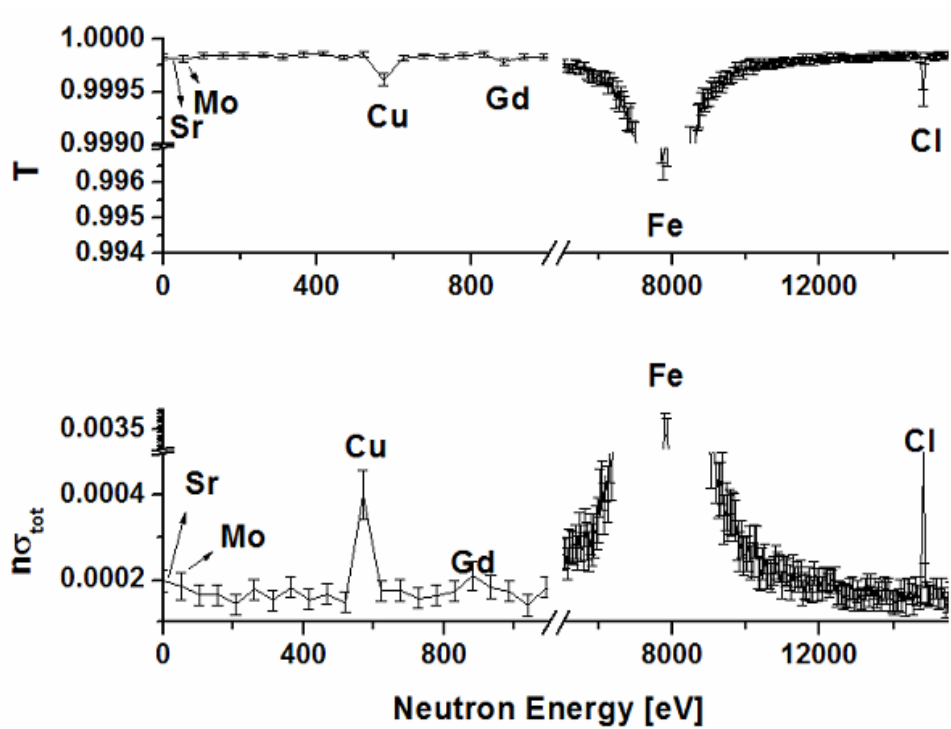

Figure 3. Simulated spectra - up); NT - down) Total cross section pondered by concentration

Similarly with the case of concentration determination 1024 simulated data in a wide energy interval were generated. The precision of measurements depends on the presence of other resonances in the vicinity of $^{160}$Gd nucleus, namely on the concentrations of others elements. The error can be improved by taking into account a larger number of experimental data or reducing the neutron energy interval to an appropriate analysis.

Due to the high resolution of NT experiments for slow neutrons even for heavy nuclei where we have many neutron resonances it is possible in principle to find a range where the resonances are isolated. One of the problem which can arise during the measurements (as in NAA or others analytical techniques) is the overlapping of the resonances from different nuclei or different isotopes of the same nucleus. This situation in principle can be solved by the measurements of partial reaction cross section as suggested in Figure 3.

Still the present computer simulation should to be improved for some strict situations. For example in the case of close resonances from different elements the precision of experiment can decrease seriously as in the interference of close resonances from the same sample. In principle such cases could be in principle avoided by searching in the spectra a more isolated resonance of the intended element as possible. From here results the fact that in future simulated experiments it will be desired to introduce another resonances as well.

# 6. CONCLUSIONS

The present simulated NT experiment for the identification of the $^{160}$Gd in an environmental sample has demonstrated the possibility of the realization of such a real neutron experiment.

The present work is an extension of light elements determination by a combination of INAA and NT realized previously by the authors [2]. As in the present work in reference [2] the authors have realized also a simulated computer experiment for the evaluation of the properties of light elements ($^6$Li and $^{10}$B) which cannot be determined by NAA methods. For the mentioned nuclei the (n,α) cross sections are very high, properties that can be used were used in NT experiments. In the case of $^{160}$Gd nucleus determination of the value of total cross section for the S resonance with energy $E_S$ = 904.6 eV is about 2800 b. This resonance is easy to be observed in NT spectra for the existing neutron pulsed source in the present. In both cases the properties of analyzed nuclei were used.

One main conclusion is that the combination of some nuclear analytical techniques based on elements properties is a proper solution in the analytical elemental analysis. This work is a piece of the nuclear data program at IREN – FLNP, JINR, Dubna.

# Acknowledgements

This study was supported by the Cooperation Program in 2013 between JINR and Romanian Research Institutes leaded by the Prof N. V. Zamfir, the Romanian Plenipotentiary Representative to the JINR Dubna.

# REFERENCES

[1] G. STEINHAUSER, S. MERZ, J.H. STERBA, J Radioanal Nucl Chem, 296, 1, 165 (2012)
[2] C. OPREA, A. I. OPREA, Proceedings of XX$^{th}$ International Seminar of Interaction of Neutrons with Nuclei, Alushta, Crimeea, Ukraine, May 21-26, ISBN 978-5-9530-0352-0, ISINN-20, 199-204, (2012)
[3] S.F. MUGHABGHAB, M. DIVADEENAM, N.E. HOLDEN, Neutron Cross Sections, NY, Academic Press, 1981, V.1
[4] G. AUDI, A.H. WAPSTRA, Nucl. Phys. A, 565, 1-65 (1993)
[5] G. AUDI, A.H. WAPSTRA, Nucl. Phys. A, 595, p. 408-490 (1993)
[6] K.J.R. ROSMAN, P.D.P. TAYLOR, Pure Appl. Chem., 71, p 1593-1607 (1999)
[7] A. FODERARO, The Elements of Neutron Interaction Theory, The MIT Press, Cambridge, Massachusetts and London, England, ISBN 0 262 06033 7 (1971)
[8] L.B. PIKELNER, Neutron Spectroscopy, VIII-th Summer School on Neutron Physics, 30 August – 9 September 1998, Dubna, Proceedings of the School, p.288, JINR Dubna Publishing Department (1999) (in Russian)
[9] L.B. PIKELNER, Neutron Sources and Spectrometers, Landolt – Bornstein Series, Elementary Particles, Nuclei and Atoms, Group1, Vol. 16, p. 4-1, Springer-Verlag, Heidelberg (2000)
[10] A.V. YGNATYUK, YU.P. POPOV, Neutron Induced Reactions, Vol. 16, p. 4-1, Springer-Verlag, Heidelberg (2000)
[11] K. SEIDEL, D. SELIGER, A. MEISTER, Z. MITTAG, V. PILTZ, Fizika Elementarnyh Chastits i Atomnogo Yadra, v. 19, 2, p.307 – 345 (1988) (in Russian)
[12] W. FURMAN, New pulsed neutron source of the JINR - the IREN project, In Proc. of Int. Conference "Nuclear Data for Science and Technology", Ed. by G.Reffo, A.Ventura and C.Grandi, Trieste, May 19-24 1997, vol. 59 part 1, p. 421-425, SIF, Bologna (1997)